# Spin-current vortices in current-perpendicular-to-plane nanoconstricted spin-valves

N. Strelkov[1,2], A. Vedyayev[1,2], N. Ryzhanova[1,2], D. Gusakova[1,a], L. D. Buda-Prejbeanu[1], M. Chshiev[1], S. Amara[1], N. de Mestier[1], C. Baraduc[1], B. Dieny[1]

[1]*SPINTEC, UMR CEA/CNRS/UJF-Grenoble1/Grenoble-INP, INAC, 38054 Grenoble, France*

[2]*Lomonosov University, Faculty of Physics, Department of Magnetism, Moscow, Russia*

**ABSTRACT**

The charge and spin diffusion equations taking into account spin-flip and spin-transfer torque were numerically solved using a finite element method in complex non-collinear geometry with strongly inhomogeneous current flow. As an illustration, spin-dependent transport through a non-magnetic nanoconstriction separating two magnetic layers was investigated. Unexpected results such as vortices of spin-currents in the vicinity of the nanoconstriction were obtained. The angular variations of magnetoresistance and spin-transfer torque are strongly influenced by the structure geometry.

Since the discovery of Giant Magnetoresistance (GMR) in 1988[1], the field of spin electronics has steadily expanded, stimulated by both fundamental breakthrough discoveries (tunnel magnetoresistance (TMR) at room temperature[2,3], spin transfer torque[4,5] (STT), voltage controlled magnetic devices[6]) and a strong synergy between basic research and industrial developments (magnetoresistive heads for hard disk drives[7], Magnetic Random Access Memories (MRAM)[8], logic devices[9], RF oscillators[10]). Several theories were proposed to explain the essence of the observed spintronic phenomena. The GMR was explained in terms of interplay of spin-dependent scattering phenomena taking place at the interfaces and/or in the bulk of neighboring magnetic layers[1,7,11]. In particular the concept of spin accumulation and spin diffusion length were successfully introduced to describe the diffusive transport in current-perpendicular-to-plane (CPP) metallic multilayers. These concepts initially developed in collinear magnetic geometry[12] have been subsequently generalized to non-collinear case[13-16]. At the same time TMR was first explained by simple quantum mechanical tunneling of spin-polarized electrons[17,18]. Later on, another mechanism of spin-filtering through crystalline tunnel barrier was proposed based on the symmetry of the electron wave-functions in the magnetic electrodes and barrier[19]. Finally STT was predicted to result from exchange interaction between spin polarized conduction electrons and those responsible for the local magnetization[4,5]. However, all these theoretical models have been applied so far only for very simple geometries with homogeneous current flow. In contrast, most spintronic devices under research or development such as point contacts[20,21], low resistance tunnel junctions[22] or GMR CPP magnetoresistive heads[7] with current crowding effects, and current confined path (CCP) structures[23,24] involve inhomogeneous current flows.

The purpose of the present numerical study was to investigate the peculiar effects which may arise in spin-dependent transport when the charge current flow is highly non-uniform for geometrical reasons. To illustrate this point, we focused this study on the case of nanoconstricted spin-valves, i.e. structures formed of two extended magnetic layers separated by a non-magnetic nanoconstriction. Using a finite element solver, we fully calculated the spatial dependence of the spin accumulation vector, charge current vector, spin current tensor, in-plane and perpendicular components of the spin-transfer torque as a function of the angle between the magnetizations of the two layers. This study illustrates that unexpected phenomena such as vortices of spin current may appear as a result of the system geometry and associated current non-uniformity. These phenomena can strongly influence the magnetization dynamics and must be properly taken into account when designing spintronic devices.

The formalism that we used was proposed by Zhang *et al*[13] and is based on a generalization of Valet and Fert theory[12] in the diffusive limit. Each material constituting the system of arbitrary shape and composition is described by local transport parameters ($C_0$–conductivity, $\beta$–spin asymmetry of $C_0$, $D_0$–diffusion constant related to $C_0$ via Einstein relation[13], $\beta'$–spin asymmetry of $D_0$, $N_0$–density of state at Fermi level).

For this study, we assumed $\beta=\beta'$. Furthermore, we only took into account bulk spin-dependent scattering. In the present approach, interfacial scattering could also be introduced by describing each interface as a thin layer having bulk properties matching the interfacial spin-dependent scattering properties[12,25]. Taking into account interfacial scattering would not change the qualitative description of the phenomena presented in this paper.

All transport properties are then described by 4 local variables: the scalar electrostatic potential $\tilde{\varphi}$ and the 3 components of spin accumulation in spin space $(m_x, m_y, m_z)$. The local charge current vector is then given by:

$$\mathbf{j}^e = 2C_0 \nabla \tilde{\varphi} - \frac{2\beta C_0}{eN_0}(\mathbf{u}_M \cdot \nabla \mathbf{m}) \qquad (1),$$

and the spin current is described by a tensor with 3 coordinates for both spin and real space as:

$$\mathbf{j}^m = \frac{2\hbar\beta C_0}{e}(\mathbf{u}_M \otimes \nabla\tilde{\varphi}) - \frac{2C_0}{e^2 N_0}\nabla\mathbf{m} \qquad (2),$$

where $\mathbf{u}_M$ and $e$ represent a unit vector parallel to the local magnetization and electron charge, $\hbar$ and $\mu_B$ are Planck constant and Bohr magneton.

The 4 variables are then calculated in steady state everywhere in space by solving the set of fundamental equations of spin-dependent diffusive transport:

$$\begin{cases} div\mathbf{j}^e = 0 & (3) \\ div\mathbf{j}^m + \frac{J_{sd}}{\hbar}(\mathbf{m}\times\mathbf{u}_M) + \frac{\mathbf{m}}{\tau_{sf}} = 0 & (4) \end{cases}$$

where $J_{sd}$ and $\tau_{sf}$ represent s-d exchange interaction constant and spin relaxation time, respectively. Eq. (3) expresses the conservation of charge while Eq. (4) states that the spin polarization of the current is not conserved. It can vary either due to spin relaxation or local spin-transfer torque given by $\mathbf{T} = \frac{J_{sd}}{\hbar}(\mathbf{m}\times\mathbf{u}_M)$. The constant $J_{sd}$ and time $\tau_{sf}$ are related to spin-reorientation length $\lambda_J = \sqrt{2\hbar D_0 / J_{sd}}$ and spin-diffusion length $l_{sf} = \sqrt{(1-\beta^2)2D_0\tau_{sf}}$, respectively[13].

Using this formalism, the spin-dependent transport was investigated in the two dimensional nanoconstricted spin-valve represented in Fig. 1. It consists of two 3nm thick magnetic layers ($M_1$,$M_2$) separated by a non-magnetic metallic nanoconstriction of 2nm thick and variable diameter. The nanoconstriction acts as a non-magnetic conducting pinhole connecting the two magnetic metallic electrodes across an insulating spacer. This central magnetic system is sandwiched between two 400nm thick non-magnetic metallic electrodes. We assume that the relative orientation of the magnetizations in the two magnetic layers can be varied in the plane perpendicular to x–axis. Voltage of $\varphi_{in} = 0V$ ($\varphi_{out} = 50mV$) is uniformly applied on the left (right) surface of the left (right) electrode, respectively. At these boundaries current flow is perpendicular to the surface whereas it is tangent to the other edges.

Using a finite element technique, we solved the system of equations 1–4 and obtained the spatial distribution of the spin accumulation and charge current vectors, spin current tensor, in-plane and perpendicular components of the spin-transfer torque as a function of the angle between the magnetizations of the two layers.

We used the following bulk parameters to represent the various materials of the system[26]: $C_0=0.005\Omega^{-1}nm^{-1}$, $\beta=0.6$, $l_{sf}=20nm$, $\lambda_J=1nm$, $D_0=1.7 \cdot 10^{15}nm^2/s$ for magnetic layers and $C_0=0.02\Omega^{-1}nm^{-1}$, $l_{sf}=50nm$, $D_0=6.9 \cdot 10^{15}nm^2/s$ for outer electrodes and nanoconstriction. These bulk parameters are representative for Co and Cu, respectively[25]. The density of states value $N_0$ corresponds to the Fermi level of Co close to 7eV. Under these assumptions, the resistance of the stack with continuous Cu spacer per 1nm of depth in parallel (antiparallel) magnetic configuration is $R_P=209\Omega$ ($R_{AP}=210\Omega$), yielding a magnetoresistance ratio $(R_{AP}-R_P)/R_P=0.5\%$.

Fig.2 shows the charge current (arrows) and electrostatic potential distribution (color map) throughout the structure in antiparallel magnetic configuration. As expected, the current converges towards the constriction and diverges afterwards, the voltage gradient being maximum within the constriction. It is interesting to note at this point that due to the convergence (divergence) of the current towards (away from) the constriction, a significant in-plane component of the charge current exists within the magnetic layers.

In Fig. 3 we present the spin current distribution of the component parallel to the *y*-axis (arrows), i.e. parallel to the magnetization of the reference layer (the layer on the left of the constriction), and the corresponding spin accumulation component (color map). Figs. 3(a)-(c) respectively correspond to parallel, perpendicular and antiparallel configurations of the magnetizations of the two ferromagnetic layers.

In parallel configuration (Fig.3(a)), the spatial distribution of the *y*-component of spin current looks very similar to the charge current distribution (Fig.2). Its amplitude gradually increases towards the constriction due to the increase of both current polarization (over length scale $l_{sf}$) and charge current density. A symmetric decrease occurs on the other side of the constriction. In this symmetric structure, the spin accumulation is zero in the middle of the constriction. On the left side of the constriction, there is an excess of spins antiparallel to magnetization (due to spin accumulation at the interface between the left electrode and reference layer) and an excess of spins parallel to magnetization on the right side.

At 90° orientation (Fig.3(b)), the *y*-component of spin current drops rapidly to zero when the electrons penetrate into the right magnetic layer. This is due to a reorientation of the electron polarization which takes place over the length scale $\lambda_J$ (~1nm) much shorter than $l_{sf}$ (~50nm). This explains why the gradient of *y*-component of spin current is much steeper on the right than on the left of the constriction.

The situation of antiparallel alignment (Fig.3(c)) is particularly interesting because it unexpectedly reveals the formation of spin current vortices on both sides of the constriction. This vorticity can be understood according to the following picture. As it was pointed out previously, due to the convergence (divergence) of the charge current towards (away from) the constriction, a significant component of the current flows in the plane perpendicular to *x*-axis within the magnetic layers on both sides of the constriction and acquires a quite large spin polarization. Thus positive spins (pointed in positive *y*-direction) travels towards the constriction on the left side in the plane perpendicular to *x*-axis and negative spins flows away from constriction on the right side in the plane perpendicular to *x*-axis. The latter is also equivalent to the convergence of positive spins towards constriction on the right side. As a result, a large flow of electrons with spins aligned in positive y-direction converges towards the constriction from both sides, yielding a very intense spin accumulation within the constriction. Then, due to the fact that spin accumulation is very large inside the constriction and rapidly vanishes away from it, additional diverging flow of positive spins away from the constriction appears on both sides along *x*-axis. The combination of the converging spin current flowing perpendicular to *x*-axis with the diverging spin current flowing along *x*-axis gives raise to spin current vortices on both sides of the constriction. These vortices are better visualized in the inset of Fig.3(c) where the *y*-component of spin current lines are plotted with uniform density.

We also computed the dependence of the CPP resistance (CPP-R($\theta$)) of the structure as a function of angle $\theta$ between the magnetizations of ferromagnetic layers (Fig.4). Two cases are compared: the CPP-R(θ) in presence of a constriction of 5nm diameter and without constriction (i.e. the constriction is replaced by a continuous Cu spacer). The presence of the constriction clearly affects the shape of the CPP-R($\theta$) variation. Interestingly both variations can be very well fitted with the expression proposed by Slonczewski in the frame of transport model combining ballistic and diffusive features [27]: $r = (1 - \cos^2\theta/2)(1 + \chi\cos^2\theta/2)^{-1}$, where *r* is the reduced resistance defined by $r = (R(\theta) - R(0))(R(\pi) - R(0))^{-1}$. The $\chi$ values in the above expression, however, are quite different in the two situations being equal to *15.86* and *4.24* for the continuous spacer and nanoconstriction, respectively. This result points out that the device geometry can strongly impact the angular variation of CPP-GMR, an effect certainly important to take into account in the design of CPP-GMR devices, particularly GMR heads for hard disk drives.

As a further step, we calculated the STT exerted by the spin polarized current on the right magnetic layer as a function of the angle between the two magnetizations (Fig.5). In the general case, STT has two components: a component in the plane formed by the magnetizations of the two magnetic layers (sometimes called Slonczewski's term[4]) and a component perpendicular to it (also called field-like term[13]). In metallic CPP spin-valves, it is generally argued that the field-like term is weak as a result of averaging over all incidences of conduction electrons penetrating in the ferromagnetic layer[15] In contrast, in magnetic tunnel junctions, the perpendicular torque amplitude can represent up to 30% of the in-plane torque value as a result of the strong decrease of transmission probability through the tunnel barrier when the incident electron momentum departs from the normal to the barrier[28].

Fig.5 (a) and (b) show the angular variation of the two components of STT integrated over the whole volume of the free layer assuming various diameter of the constriction (2nm, 5nm or continuous spacer). Our results confirm that in this diffusive approach, the perpendicular component of STT is two orders of magnitude lower than the in-plane torque. It is interesting to note that the shape of the angular variation of STT is quite similar for the two components. Actually, these shapes can also be very well described by the expression proposed by Slonczewski for the reduced torque[27]: $\tau(\theta) = \sin\theta(\Lambda\cos^2(\theta/2) + \Lambda^{-1}\sin^2(\theta/2))^{-1}$ both for the in-plane and perpendicular components. However, as for the angular variation of GMR, the $\Lambda$ fitting parameter strongly depends on the constriction diameter ($\Lambda$ respectively equals to 1.75, 2.19, 4.09 for 2nm, 5nm and continuous spacer). Note that the equality[27] $\Lambda^2 = \chi + 1$ is verified quite well for the case of laterally homogenous electron current, i.e. in the case of the continuous spacer. It should be emphasized that the formula for reduced torque was obtained in the case of "standard" (or "symmetric") metallic structure where both magnetic layers have close physical parameters (for example, Co/Cu/Co type structures). This is also the case in our numerical study. We find that the agreement of our calculations with Slonczewski's formula is quite reasonable in this case. The diffusive scattering[13] does not modify the form of Slonczewski expression but is indirectly hidden in the $\Lambda$ parameter. In the case of strongly asymmetric structures ( for example so-called "wavy" structures[29]) one should rather use more general expression for reduced torque proposed in Ref.30.

Fig.5(c) shows a map of the in-plane STT amplitude for 90° magnetic orientation in the case of a constriction of 10nm diameter. Clearly, the STT is most important in the immediate vicinity of the constriction where the current density is the largest. Actually the gradient of STT is quite large since the charge current density drops very quickly around the edges of the constriction. That points out that new length scales may emerge in these confined geometries due to a balance between spin torque gradient and exchange stiffness. In micromagnetic simulations, traditionally only two length scales are considered: the Bloch wall width (balance between anisotropy and exchange stiffness) and exchange length (balance between magnetostatic energy and exchange stiffness). It is likely that additional length scales will have to be considered in structures wherein strong current gradients are imposed by the system geometry. Such new length scales imposed by the system geometry has already been introduced for instance in the context of domain walls confined in magnetic nanoconstrictions[31].

In conclusion, a finite element numerical approach has been developed to compute the charge and spin current in magnetic structures of arbitrary shape and composition. The case of 2D nanoconstricted symmetric spin-valves was treated as an illustration. Charge and spin current clearly behave very differently as demonstrated for instance by the formation of spin current vortices. The approach can be straightforwardly extended at three dimension and taking into account interfacial scattering. This type of approach could be helpful in the design of functional spintronic devices as well as for the quantitative interpretation of experimental data in devices with non uniform or non-local currents such as lateral spin-valves[32].

This project has been supported in parts by the European RTN "Spinswitch" MRTN-CT-2006-035327, the ERC Adv grant HYMAGINE and by Chair of Excellence Program of the Nanosciences Foundation (Grenoble, France). NS, AV and NR are grateful to RFBR for partial financial support.

FIGURES CAPTIONS

FIG. 1. (color online) Model system used for the finite element calculation of CPP spin transport through a nanoconstricted spin-valve. The metallic pinhole (PH) connecting the two ferromagnetic layers is 2 nm thick and of variable diameter. Note, that in the case of 2D model all quantities are calculated per 1nm depth. Thus the cross-section surface of the stack with continuous Cu spacer is 100nm×1nm and the volume of the free magnetic layer is 3nm×100nm×1nm.

FIG. 2. (color online) Zoom around the nanoconstriction showing the charge current flow (arrows) through the constriction and electrostatic potential (color mapping) corresponding to the antiparallel state.

FIG. 3. (color online) Zoom around the nanoconstriction: *y*-component of spin current (black arrows) and *y*-component of spin accumulation (color mapping) for three magnetic configurations: (a) parallel, (b) 90°, (c) antiparallel. In Fig.3(c), the white arrows remind the charge flow and the grey closed arrows indicate the formation of spin current vortices which are better evidenced in the inset.

FIG. 4. Angular variation of the CPP reduced resistance for the constriction of 5 nm diameter and continuous spacer. The dots are the calculated values and the lines are fits according to Slonczewski's expression (see text).

FIG. 5. (color online) (a) In-plane and (b) perpendicular components of averaged spin-transfer torque over the whole volume of the "free" (right) magnetic layer as a function of the angle

between the magnetizations. (c) Mapping of the amplitude of the in-plane torque for 90° magnetic configuration in the presence of nanoconstriction with 10 nm diameter.

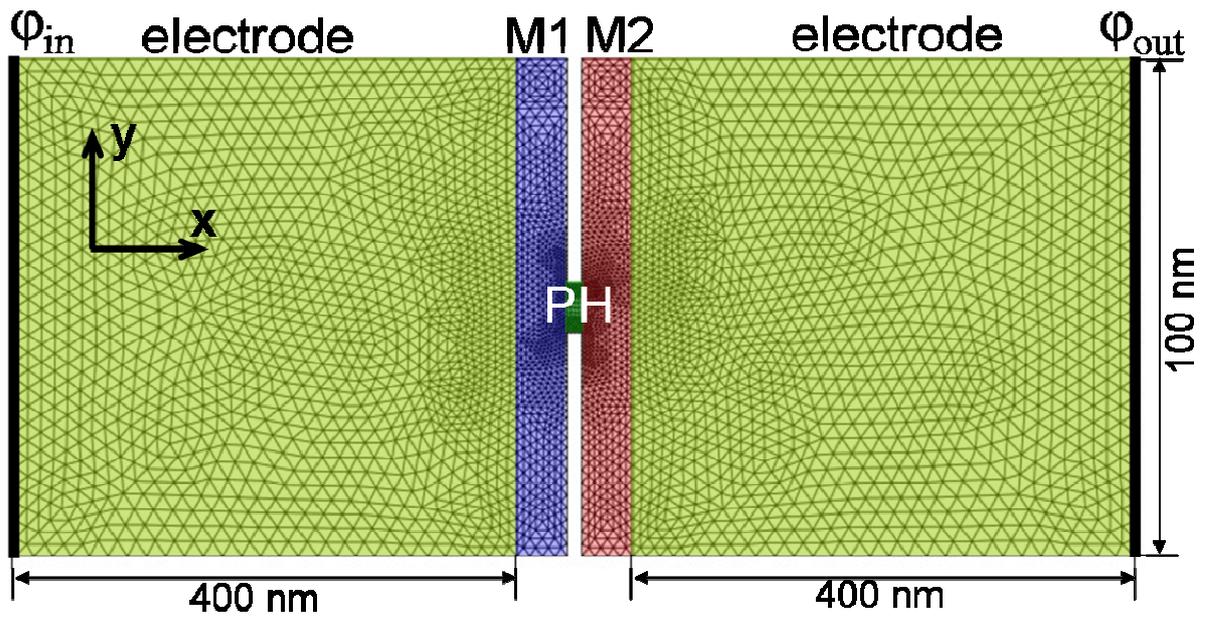

FIG. 1. N. Strelkov *et al*

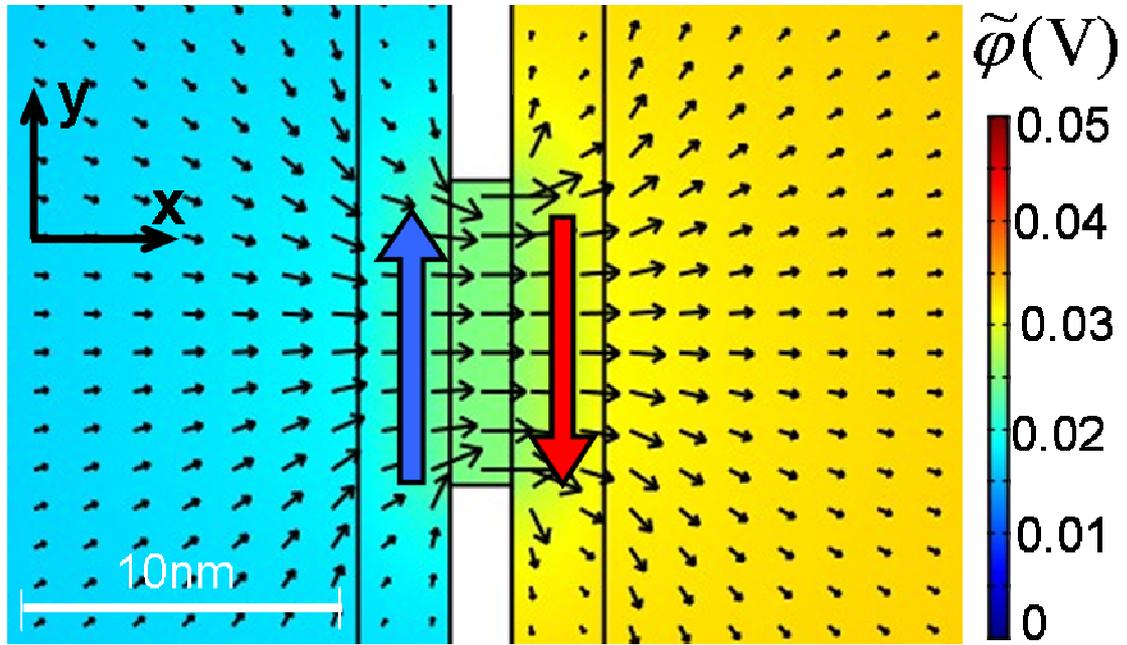

FIG. 2.                                                                                     N. Strelkov *et al*

FIG. 3. N. Strelkov *et al*

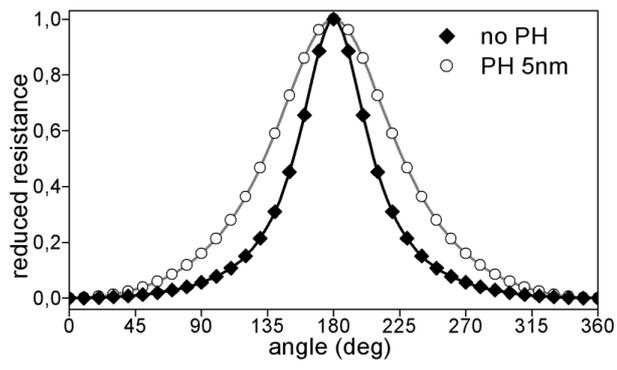

FIG. 4. N. Strelkov *et al*

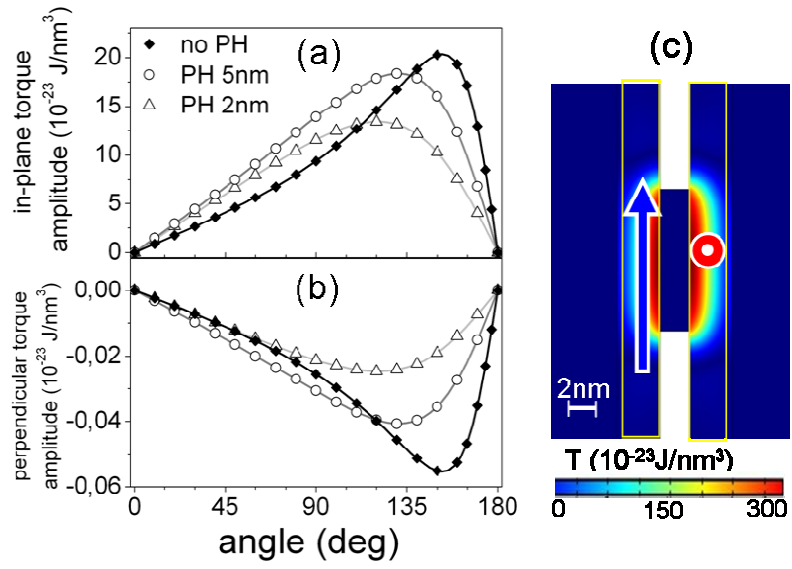

FIG. 5. N. Strelkov *et al*